# Phase imaging by spatial wavefront sampling

F. SOLDEVILA,[1,*] V. DURÁN,[2,3] P. CLEMENTE,[1,4] J. LANCIS,[1] E. TAJAHUERCE[1]

[1]GROC·UJI, Institute of New Imaging Technologies (INIT), Universitat Jaume I, E12071 Castelló, Spain.
[2]Univ. Grenoble Alpes, LIPHY, F-38000 Grenoble, France.
[3]CNRS, LIPHY, F-38000 Grenoble, France.
[4]Servei Central d'Instrumentació Científica (SCIC), Universitat Jaume I, E12071 Castelló, Spain.
*Corresponding author: fsoldevi@uji.es

Phase imaging techniques extract the optical path-length information of a scene, whereas wavefront sensors provide the shape of an optical wavefront. Since these two applications have different technical requirements, they have developed their own specific technology. Here we show how to perform phase imaging combining wavefront sampling using a reconfigurable spatial light modulator with a beam position detector. The result is a time-multiplexed detection scheme, capable of being shortened considerably by compressive sensing. This robust reference-less method does not require the phase unwrapping algorithms demanded by conventional interferometry, and its lenslet-free nature removes tradeoffs usually found in Shack-Hartmann sensors.

## 1. Introduction

Even though the physical nature of light has been fully understood for more than a century, there are still no available detectors capable of directly imaging both the amplitude and phase information of a wavefront (Fig. 1a). Information about those two quantities is of capital interest when trying to perform biomedical imaging [1,2], aberration measurement and correction in visual optics [3,4], and three-dimensional imaging [5], among other applications. The limitation in performing optical measurements ultimately arises from the extremely fast oscillations of the optical fields, which current electronics are unable to resolve. As detectors only capture light irradiance, several approaches have been proposed over time to tackle the phase problem. Historically, Gabor suggested in 1949 the first quantitative technique, which used interferometric information to recover the complex optical field [6], stablishing the basis of modern holography and paving the way for applications such as quantitative phase microscopy [7]. In parallel, phase information plays a fundamental role in adaptive optics, a technique that intends to measure and correct optical aberrations in real time [8]. Although adaptive optics was initially conceived for circumventing atmospheric turbulences in Astronomy, its simple operation principle has found applications in other areas, such as visual optics [3,4], microscopy [9,10], and biomedical imaging [11,12].

Two main groups of techniques have emerged to tackle the problem of recovering the complex amplitude of the light field. The first one includes interferometric approaches, which measure the interference between light coming from the object and a reference beam (Fig. 1b). Although they are extremely powerful for conducting precise phase measurements, their high sensitivity to environmental perturbations (such as mechanical vibrations and changes in temperature) and the need of a reference beam (not always attainable) hinder the implementation of portable and compact interferometric imaging systems in many applications. As an alternative, a second group of techniques has emerged, whose objective is to recover the same information without the need of a reference beam. These non-interferometric approaches rely on several assumptions about the object beam, and use mathematical algorithms to infer the wavefront information [13]. Examples of this kind of techniques are Fourier ptychography [14], coherent diffractive imaging [15,16], phase imaging based on the transport-of-intensity-equation [17], and phase imaging with randomized illumination [18]. By eliminating the need of a reference beam, simpler and more robust devices can be designed. Furthermore, computational approaches also enable to extend the physical capabilities of imaging systems, providing increased field of view [19], optical sectioning [20] or superresolution [21,22]. However, the recovery algorithms used in those techniques usually entail high post-processing times or multiple data acquisitions to recover one image.

Among the non-interferometric techniques, Shack-Hartmann (SH) wavefront sensors [23] are currently the most employed for measuring optical aberrations. SH wavefront sensors combine a lenslet array with a pixelated detector, like a CCD or a CMOS camera. By placing the detector at the focal plane of the lenslet array, the positions of the foci generated by each lenslet can be linked to the phase of the wavefront in each region (Fig. 1c). In other words, each lenslet samples the phase information of a region of the wavefront. This method provides a very compact system, easy to use and align, and with no complex post-processing stages. For these reasons, SH sensors are used in manifold scientific fields [4,24–27]. However, the number of lenslets, their focal length, and their diameter limit the spatial resolution, dynamic range, and sensibility of SH wavefront sensing. Although multiple solutions have been proposed to manage the several tradeoffs between the above magnitudes [28–36], manufacturing processes still place a boundary on the size and curvature of available lenslets, constraining the attainable spatial resolution of commercial SH sensors. Quadriwave lateral shearing interferometry [37], also known as a modified Hartmann mask technique, makes it possible to increase the SH spatial resolution by a factor four [38]. Alternatively, phase imaging can be performed using pyramid wavefront sensors, which employs a four-sided refractive pyramid to split the Fourier plane in four parts, each one generating a phase gradient image on a pixelated detector [39]. Further developments replace the pyramid by a quadri-foil lens [40] or a liquid crystal display [41]. Despite the benefits of this approach in terms of spatial resolution, the illumination numerical aperture determines the detectable dynamic range, so in practice only relatively smooth phase gradients can be precisely recorded [42].

Here, we perform phase imaging using an alternative non-interferometric approach to measure the complex amplitude of a wavefront. We overcome the inherent limitations in spatial resolution, optical efficiency and dynamic range that are found in SH wavefront sensing. We sample the wavefront by using a set of binary amplitude masks generated by a high-speed spatial light modulator. A single focusing lens forms a time-dependent light distribution on its focal plane, where an irradiance detector is placed. Measuring the changes of the total irradiance and the centroid position of that distribution, both the amplitude and phase of the wavefront can be recovered (Fig. 1d). One advantage of this time-multiplexed detection scheme is that it can be applied using compressive sensing (CS), allowing us to go beyond the Shannon-Nyquist condition, with the subsequent reduction in acquisition time [43,44]. Unlike previous phase retrieval methods based on SH wavefront sensors or wavefront modulation [45–47], our approach is lenslet-free and does not rely on any kind of iterative algorithms. Moreover, the use of phase unwrapping algorithms is not required, such as those employed in interferometric approaches due to their limited phase unambiguity range [48]. As a proof of concept, we perform aberration sensing, as well as phase imaging of a low-contrast sample with variable optical thickness. The corresponding results are compared, respectively, with Shack-Hartmann wavefront sensing and phase-shifting interferometry.

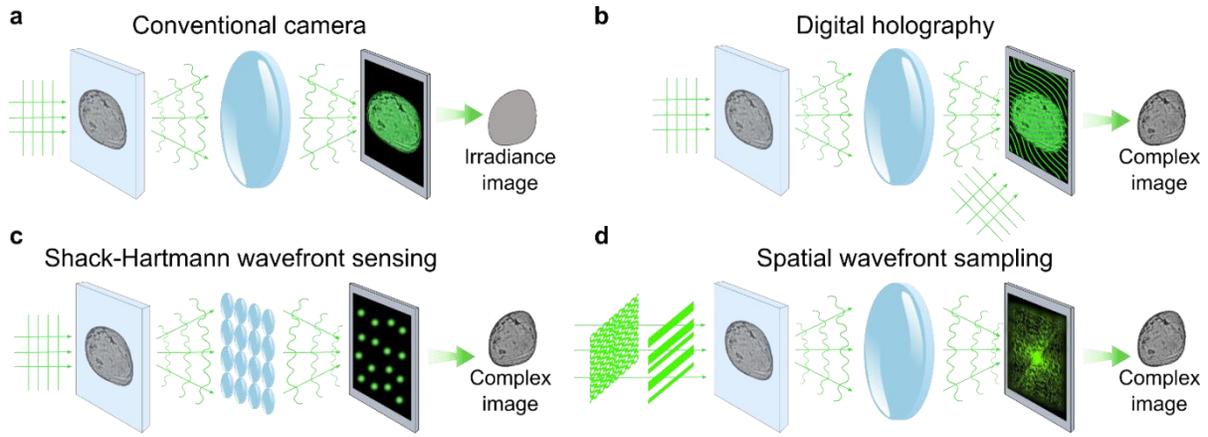

Fig. 1. Complex amplitude retrieval using spatial wavefront sampling compared to current imaging and wavefront sensing techniques. (a) In a conventional imaging system, the phase information carried by the light coming from an object (for example, a biological sample) is completely lost, as the object image is formed onto a detector that simply measures the light irradiance. (b) Digital holography captures the interference between the light coming from the object and a reference beam, allowing one to retrieve the complex amplitude of the object from irradiance measurements. (c) Shack-Hartmann sensors measure the displacements of different foci generated by an array of lenslets. From those displacements, the wavefront impinging onto the lenslet plane can be reconstructed, providing information about the phase variations introduced by the object. (d) Our technique uses a sequence of illumination amplitude patterns and a single focusing lens. The object's complex amplitude can be obtained from the statistical properties of the irradiance distribution measured in the Fourier plane of the lens. The same operation principle can be applied if, instead of the configuration shown in the figure, the light coming from an object is modulated by a set of patterns generated on a spatial light modulator.

## 2. Operation principle

The basic idea of our technique is to use the relationship between the amplitude and phase of a wavefront and the statistical moments of its Fourier irradiance distribution. This idea is also at the heart of current SH sensors, as they sample a wavefront by means of an array of lenslets and measure the centroid position (the first-order statistical moment) of the Fourier distribution created by each lenslet, allowing one to derive the local slope of the wavefront. By placing a pixelated detector (typically a CCD) at the focal plane of the lenslet array, SH sensors fully map the local slopes of the wavefront. After that measurement, numerical integration provides the phase of the wavefront at each lenslet position [49] (see Fig. 2a).

Several tradeoffs arise from the SH configuration. First, the spatial resolution of the measurement is determined by the total number of lenslets and their size. In order to increase the spatial resolution, arrays with higher density of lenslets can be built. However, reducing the size of each lenslet also decreases the amount of light at each region of the detector. If the signal-to-noise ratio of the measurement is low, the accuracy of the centroid position detection is greatly reduced, which hinders a reliable wavefront reconstruction. Even if the detector can work in low-light-level scenarios, building large arrays of lenslets with a diameter below a hundred of microns and an accurate curvature is technologically challenging [50]. Moreover, centroid displacement and wavefront slope are related via the lenslet focal length, so, for a given slope, the higher the focal length is, the larger the centroid displacements that are measured. This brings about a second tradeoff, since a strongly aberrated wavefront may produce centroid displacements larger than the size of the detection area allocated to every lenslet on the sensor. The result is a crosstalk between nearby spots that produces errors in the reconstructed phase, limiting the attainable dynamic range of the sensor. This problem can be circumvented (without sacrificing sensitivity) by a number of techniques that track the real spot location or infer it by using computational techniques [29]. Instead of trying to expand the dynamic range of a sensor by increasing its hardware and/or software complexity, an alternative approach is to implement a reconfigurable (adaptive) SH sensor that includes an array of diffractive lenslets programmed onto a spatial light modulator [28,33]. The lenslet characteristics can be then chosen to match the requirements of a specific application optimally, albeit the tradeoff between sensitivity and dynamic range still remains. Another aspect of SH sensors that is usually not considered is the fact that the total amount of light arriving at each detector region provides a measurement of the light power at the corresponding lenslet position. Using that information, one can map the irradiance of the light coming from an object. However, due to the relatively poor spatial resolution of SH sensors (usually around a thousand of lenslets), they barely can compete with interferometric systems to measure the complex amplitude of an object with high spatial frequency content.

Another approach for measuring the local slope of a wavefront is the use of a small scanning aperture and a single lens rather than an array of lenslets (Fig. 2b). For each position of the aperture, only light from a region of the wavefront will be focused by the lens. This light will generate a focal spot on the Fourier plane of the lens, where the detector is placed. For a small enough aperture, the wavefront can be locally approximated by a plane wave, and it can be demonstrated (see Methods) that the position of the centroid of the distribution is related to the gradient of the phase inside the aperture by the equation

$$\vec{\Delta} = (\Delta x, \Delta y) = \frac{\lambda f}{2\pi} \vec{\nabla}\varphi, \quad (1)$$

where $\vec{\nabla} = \left(\partial/\partial_x \hat{i}, \partial/\partial_y \hat{j}\right)$ represents the gradient operator in two dimensions, $\lambda$ is the light wavelength, and $f$ is the focal length of the focusing lens. One can see the combination of this small aperture and the single lens as one of the lenslets of a SH array. By scanning the wavefront with the aperture, the phase can be easily recovered by numerical integration. Furthermore, calculating the zero-order moment of the light distribution (i.e., its total irradiance) provides a measurement of the wavefront amplitude in the position of the aperture, leading to the recovery of an amplitude image. Although this approach would solve the spatial resolution problem of SH sensors, given that it is easier to generate small apertures than small lenslets, it still presents a tradeoff between the spatial resolution and the amount of light at each area of the detector plane, unless the scanning is performed with a laser combined with a galvanometric mirror.

Our technique resolves the lack of efficiency of the scanning-based approach and makes it possible a compressed sensing process. As can be seen in Fig. 2c, instead of just selecting a small region of the wavefront, a time-variable mask (a sequence of programmed patterns) selects simultaneously several regions of the wavefront. This procedure resembles those used in structured illumination microscopy or in single-pixel imaging [44,51,52]. It must be noted that, as in single-pixel imaging, the technique presented here can work in two different configurations: either the object (or wavefront) under study can be imaged onto the spatial light modulator or the codified patterns can be projected onto the object plane [53–56]. The key factor that has to be accomplished is that both the mask and the object wavefront overlap in one plane. Even though the irradiance distribution generated by the focusing lens results from the contribution of light coming from different areas of the sampling plane, the amplitude and phase information can be still retrieved, provided that the sequence of sampling patterns is known (see Methods). Now, the amount of light arriving at the detector for every pattern has been greatly increased in comparison to the scanning-aperture scheme. This fact is commonly known as the multiplex or Fellgett's advantage, and has been extensively used to increase the signal-to-noise ratio of optical measurements over the last decades [52,57–59]. In our proposal, the set of masks used are a shifted and rescaled version of the Walsh-Hadamard functions [60], since this election provides several benefits. First, Walsh-Hadamard functions are binary, so they are a good choice when working with binary amplitude modulators, as is the case of digital micromirror devices (DMDs), which reach very high refresh rates (around 22 kHz). Second, the sampling patterns have the same number of absorbing and transmitting (or reflecting) areas, no matter the size of those areas. This fact is important, because once the window size on the modulator has been fixed, increasing or decreasing the spatial resolution of the masks does not change the total amount of light transmitted or reflected by the device. Then, irrespective of the chosen spatial resolution, each measurement always works with the same number of incident photons, unlike the wavefront sensing approaches described above. Furthermore, since in every measurement only a single spatial light distribution is detected, the cross-talk errors that limit the dynamic range in a SH sensor are not present here.

In order to recover the amplitude and phase information of a wavefront, our method requires the measurement of the zero and first order statistical moments of the spatial distribution of light, i.e., its total irradiance and the position of its centroid. The classical approach to measure those quantities is to place a pixelated sensor in the Fourier plane of the lens to acquire a digital image, and then calculate them computationally. However, other experimental approaches can be explored. Here we have used a lateral position detector instead of a camera. The benefit of using this detector lies in its capability to provide the information about the power of the beam and the position of its centroid at speeds in the order of several kHz, without need of computational procedures. Despite digital cameras offer better sensitivity and accuracy in the detection of the centroid, we have opted for a simpler design to demonstrate the feasibility of our approach.

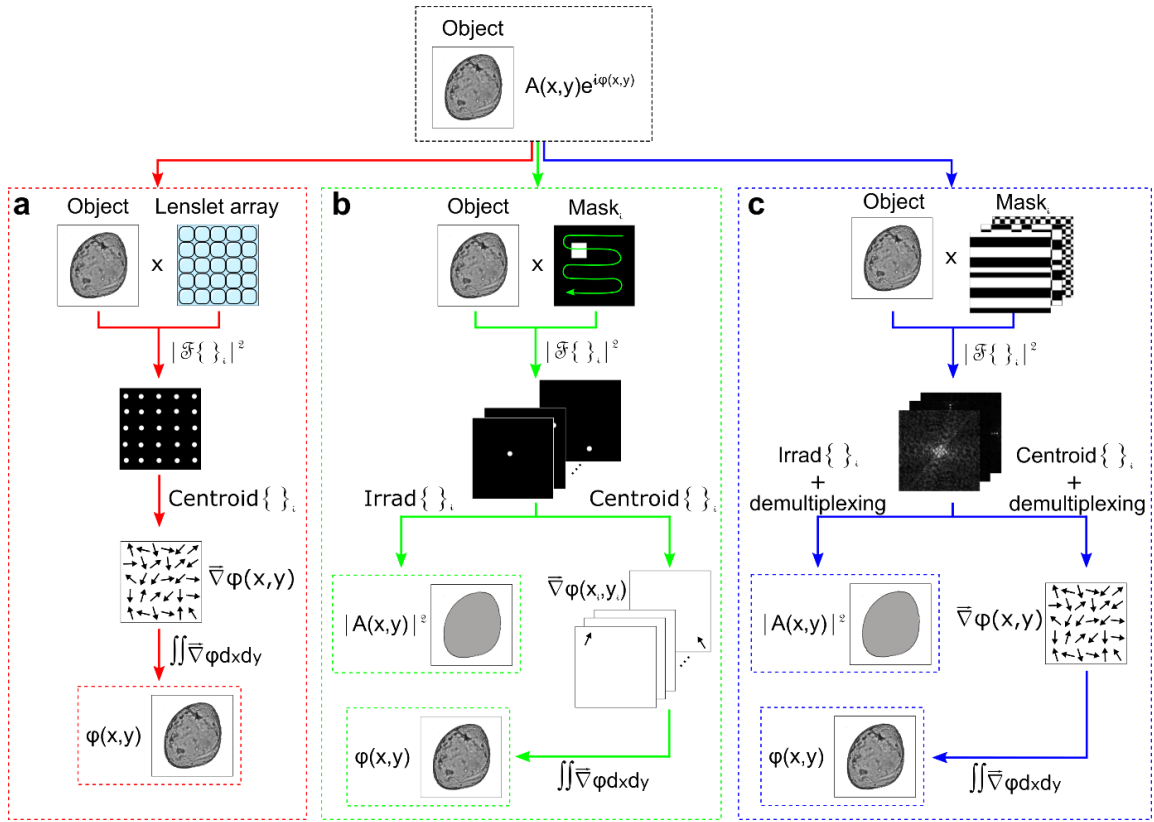

Fig. 2. Operation principle of different wavefront sensing approaches. (a) Shack-Hartmann wavefront sensing. A wavefront coming from an object passes through an array of lenslets, which produce a distribution of focal spots on the detector. The position of each spot is linked to the local slope of the wavefront on every lenslet. Numerical integration of the slope data provides the phase of the wavefront. (b) Raster scanning wavefront sensing. A small amplitude aperture is moved over the wavefront plane and, for each consecutive scanning position, the Fourier distribution of the emerging light is recorded. The total irradiance of that distribution at each aperture position can be used to recover the amplitude image of the object. Additionally, the centroid relative location of each Fourier distribution provides the local slope of the wavefront at each scanning position, which allows one to obtain the phase image of the object. (c) Spatial wavefront sampling. Instead of using a small scanning aperture, the wavefront coming from the object is sampled by a set of amplitude masks (reconstruction basis). Now, from the irradiance of each Fourier distribution generated on the detector plane, one can measure the mathematical projection of the object amplitude into the reconstruction basis. By demultiplexing that information, the object amplitude is spatially resolved. In the same way, demultiplexing the data concerning the centroid locations provides the slope map of the wavefront and then, after numerical integration, the object phase image.

## 3. Results

### A. Experimental verification

To demonstrate our method, we present the experimental device shown in Fig. 3a. Light coming from a laser (Oxxius slim-532) emitting at 532 nm is collimated with a lens (L1) and impinges onto a DMD (DLP Discovery 4100 V-7000 from ViALUX). By using a 4-f system, formed by lenses L2 and L3, light is projected onto an object, which modifies the wavefront phase. After going through it, light passes through a condensing lens (CL), with a focal length of 150 mm. In its Fourier plane, the lateral position detector (Thorlabs PDP90A) measures the irradiance of the beam and the position of its centroid. To retrieve the amplitude and phase information of the object, a full set of Hadamard patterns must be projected. The size of this set depends on the pixel size of the image one wants to obtain. Then, for an $N \times N$ image, $N^2$ patterns must be sent (see Methods). This sequential acquisition represents a limitation of our technique. Whereas a SH sensor captures all the information in a single shot, our technique relies on the sequential sampling of the wavefront to work.

As a first example, we show the results for a photoresist plate representing a typical coma aberration (Fig. 3c-d). In this case, the spatial resolution was set to $64 \times 64$ pixels, with a pixel pitch of 82.08 μm. For this size, the total acquisition time was roughly one second. This is not caused by the refresh rate of our DMD (that would need roughly 200 ms to project $64^2$ patterns), but by the limited bandwidth of our detector (Fig. 3b). Faster lateral position sensing detectors, which are commercially available, would speed up the acquisition process. A comparison of this result with those obtained by other wavefront sensing techniques (using a SH sensor and through an interferometric measurement) can be found in Supplement 1.

### B. Comparison with Shack-Hartmann wavefront sensing

To test the capabilities of the proposed technique, we have compared it with a commercial SH wavefront sensor (Thorlabs WFS150-5C) in several scenarios. As a first object, we employed a spherical lens (Fig. 4a) with a focal length of 500 mm aligned with the optical axis. This simple choice facilitates to compare both methods quantitatively, since we expect to recover the phase of a spherical wavefront. The object plane was imaged onto the lenslet array plane of the SH wavefront sensor by using a 4-f system. The focal spots produced by the lenslet array can be seen in Fig. 4b. Using the displacements of each focal spot respect to a reference position (previously determined by a calibration measurement), the 3D view of the wavefront can be recovered by direct integration. To ease the visualization, we also present a wrapped view of the wavefront phase (Fig. 4c).

Using the optical setup shown in Fig. 3a, we reconstructed the wavefront phase by using spatial wavefront sampling. A small region of the full measured 64×64 map of centroid displacements is shown in Fig. 4d. The modulo-2π phase reconstructed from those displacements can be observed in Fig. 4e. The total phase change over the aperture is 9.67λ, a value that differs in 0.06λ from that obtained using the SH sensor.

As a second experiment, we moved the lens away from the optical axis, thus producing a larger phase gradient. Given the technical specifications of our SH sensor, the maximum measurable total phase change over the aperture of the sensor is around 100λ. When the phase gradient produced by the off-axis lens produces a higher phase variation, the focal spots move so much in one direction that cross-talk appears and the sensor is not able to provide reliable results.

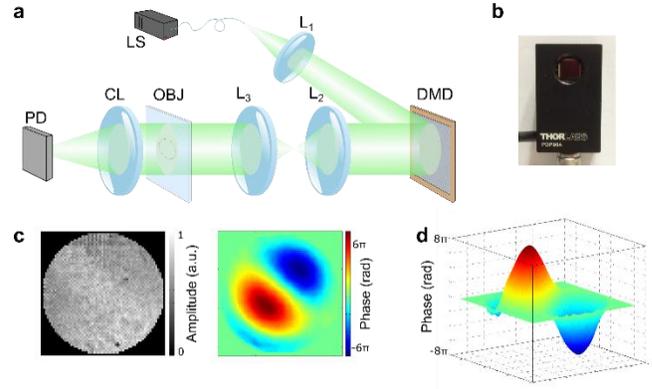

Fig. 3. Experimental verification of the proposed technique. (a) Schematics of the system. Captions: LS, laser source; $L_1$, $L_2$, and $L_3$, lenses; OBJ, object; CL, condensing lens; PD, position detector. (b) Image of the detector used in our experiments. It includes four electrodes connected to a metallic surface, whose voltage measurements provide both the irradiance of the light beam and the position of its centroid. (c) Experimental results for a plate simulating a coma aberration. As the sample is transparent, the amplitude image provides no information about the object. In the phase image, the object information is clearly recovered. (d) 3D view of the recovered phase.

In the top row of Fig. 5, we show the result obtained by using the SH sensor at the limit of its dynamic range (here we show a wrapped view of the phase to ease visualization). This limit can be considerably overcome using the setup shown in Fig. 3a. As can be seen in the bottom row of Fig. 5, spatial wavefront sampling allows us to measure a total gradient phase of 217λ. The above improvement in the dynamic range is due to the relative increase in the size of the detector area attained by our technique. In a SH sensor, the detector is divided into a number of regions equal to the number of lenslets in the array. Each region, containing a small number of pixels, is used to map the position of each focal spot and sets a limit to the maximum spot displacement. As was mentioned before, walking off beyond that limit makes several focal spots to share the same region of the detector, leading to significant errors in the reconstructed phase. Since in our technique we sequentially detect the light distribution as a whole, the full size of the sensor can be used in every consecutive measurement. As a result, the cross-talk penalty typically found in a SH sensor is not present anymore. In the same way as in the design stage of SH sensors, fine tuning of the focal length of the condenser lens used and the size of the detector will provide bigger or smaller dynamic ranges and sensitivities in the phase detection.

### C. Use of Compressive Sensing

Given the similarities between our technique and single-pixel imaging, some advanced undersampling approaches can be used to improve both the acquisition speed and the light efficiency of our method.

CS provides a recovery framework for signals sampled below the limit imposed by the Shannon-Nyquist theorem [43]. For a sequential signal sampling, this implies a reduction in measurement time. Once the signal has been sampled, an optimization algorithm provides an estimation of the original signal. This estimation is usually based on sparsity constraints and provides a high-fidelity reconstruction (see Methods).

In our experiments, we have used standard CS algorithms ($l_1$-magic package [61]) to recover the coma aberration presented in Fig. 3. The results are shown in Fig. 6.

As can be seen, for relatively smooth aberrations, estimations with fidelity higher than 90% can be recovered by using 10% of the total number of measurements established by the Shannon-Nyquist criterion. Fig. 6 includes two insets that show individual reconstructions for compression ratios corresponding to correlation coefficients over 0.9 and 0.99, respectively (points A and B in the fidelity plot). As in our commercial Shack-Hartmann sensor, the recovered wavefronts are expressed in the Zernike basis and high-resolution images are generated from that expansion (see Supplement 1 for details). By taking this into account, high-resolution acquisitions in 20 ms could be achieved operating at the full frame rate of the DMD (~20 kHz), allowing the capability to perform real time aberrometry. This could be very useful in ophtalmic scenarios, where the patient needs to stand still while the eye aberrations are measured. Moreover, even though half of the source light is not used when operating with Hadamard functions, the reduced number of projections makes the system more efficient in global terms of photon usage that a raster scan approach (even when a galvanometric mirror is employed). This could be convenient in biological scenarios, where photodamage thresholds limit the amount of light that can be used at each region of the sample. By using this CS-based procedure, one would be able to increase the measured signal level without causing photodamage when compared with the raster scan approach.

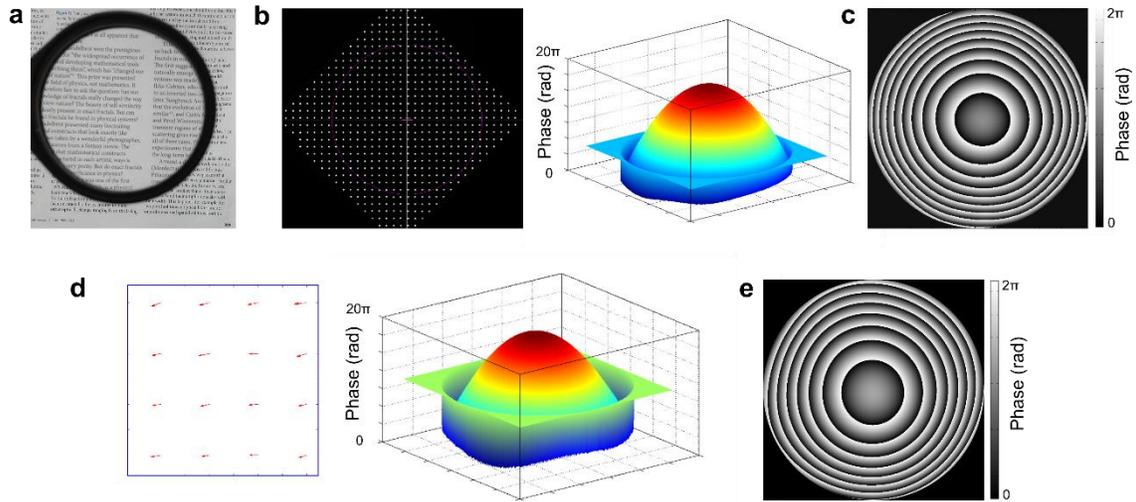

Fig. 4. Reconstruction of a spherical wavefront and comparison with a commercial Shack-Hartmann sensor. (a) Caption of the lens used as a phase object. (b) Spot map generated by the SH lenslet array. By comparing the position of each spot to a reference value, the spot displacements can be calculated. From those data, a 3D view of the phase can be recovered. (c) Wrapped phase of the lens. (d). Small region of the full displacement map for our system. Those displacements can be related to the gradient of the phase. Numerical integration of those gradients provides the phase of the object, and after decomposition in the Zernike basis, a high-resolution 3D view of the phase can be displayed (plot on the right). (e) Wrapped phase of the lens obtained using the proposed technique.

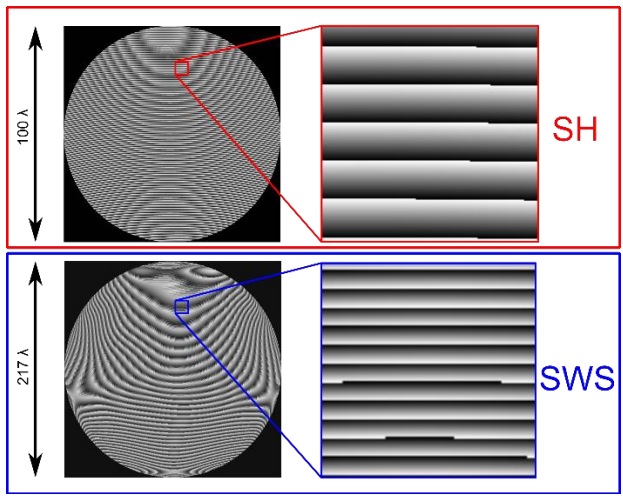

Fig. 5. High dynamic-range measurement and comparison with a commercial SH sensor. In the top row, we show the wrapped phase measured with the aid of a SH sensor when the lens shown in Fig. 4a is placed in an off-axis position. The maximum phase gradient measurable by the commercial sensor is 100λ. A zoomed region, marked in red, is shown in the right part of the figure. In the bottom row, we show the results for the same object obtained with the system shown in Fig. 3a. In this case, the lens has been displaced a bigger distance from the optical axis. The phase gradient measured is 217λ. A zoomed region, marked in blue, is shown in the right part of the figure.

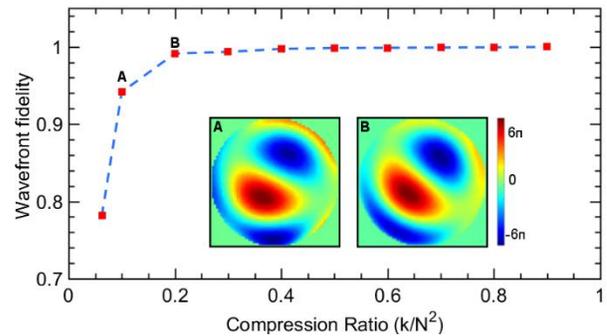

Fig. 6. Evolution of reconstruction quality when Compressing Sensing is used for wavefront recovery. For different values of the compression ratio ($k/N^2$), we present the correlation coefficient between the CS estimation and the recovery without using any undersampling. Here, the images have a resolution of $64 \times 64$ pixels, so $N^2 = 4096$. The points represent the median of the correlation coefficient for 100 different realizations of the algorithm (each realization corresponding to a different subset of random measurements).

*D. Complex amplitude retrieval and comparison with phase-shifting interferometry*

Our technique offers the possibility of reconstructing not only the phase distribution of a wavefront coming from an object but also its amplitude. For the set of illumination patterns, the light power measured by the detector provides the projections of the wavefront amplitude into the basis of Hadamard functions. Measuring all the successive projections, the object amplitude image can be recovered offline, following the operation principle of single-pixel imaging [44] (see Methods). A simple example with an object composed of an amplitude mask attached to the lens used in the previous section can be seen in Supplement 1.

In principle, the complex amplitude of a wavefront can be retrieved by means of a SH sensor, but with a relatively poor spatial resolution (given by the lenslet diameter and the lens fill factor). As a consequence, spatial features smaller than a few hundreds of microns are lost in the recovery process. Even though high resolution SH sensors are being developed and can be used to perform phase imaging, the tradeoffs between dynamic range and sensitivity are still present [62]. In our technique, the limit in the spatial resolution is ultimately given by the modulator pixel size (typically ~10 microns) and the magnification of the projecting system. In our setup, this supposes an increment in the resolution of one order of magnitude respect to a typical SH sensor.

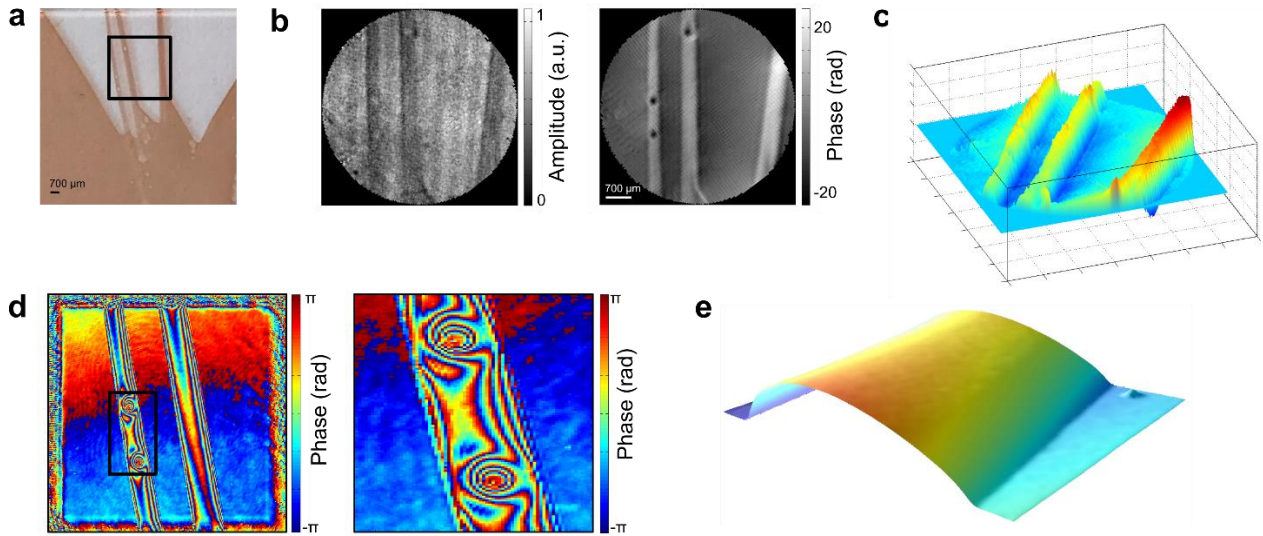

Fig. 7. Comparison with phase-shifting interferometry. (a) Photograph of the photoresist layer used for the experiment. The black square represents the region that will be imaged, consisting of zones with and without photoresist material. (b) Amplitude and phase images obtained with the proposed technique. Due to the absorption properties of the object, the amplitude image presents poor quality. However, in the phase image, fine details of the sample can be observed. (c) 3-D representation of the phase obtained. (d) Phase image obtained with phase shifting interferometry for the same region of the object. In the inset, we show the region between the two small holes present in one of the photoresist bars. (e) Physical profile of the photoresist strip between the two holes obtained with an optical profilometer (Sensofar Plµ 2300).

To demonstrate this improvement, we use as a sample a thin layer of a photoresist material. We placed the photoresist over a transparent plate, creating an object with different regions with and without material, as can be seen in Fig. 7a. Due to the difference of refractive index between the photoresist and the air, an impinging wavefront acquires a spatial phase distribution when it passes through the object. By using the setup shown in Fig. 3a, the amplitude and phase information of the sample is recovered (Fig. 7b). A 3D visualization of the phase information is shown in Fig. 7c. The images have a spatial resolution of $128 \times 128$ pixels, with a pixel pitch of $41.04$ µm. Since both the photoresist and the transparent plate present a similar absorption in the green region of the spectra, the amplitude image presents low quality. However, fine details of the sample are recovered in the phase image, such as the small holes (with a diameter of around 80 µm) present in the photoresist stripes, produced by air bubbles in the manufacturing process. These small spatial features cannot be retrieved by our SH sensor, as the diameter of each lenslet is 150 µm. Additionally, due to the low number of lenslets ($39 \times 31 = 1209$) the SH sensor would only provide an image clearly "pixelated".

In order to check the validity of our results, we resort to phase-shifting interferometry. This technique uses a pixelated detector, which makes it possible to exploit the high spatial resolution of a commercial CCDs (in our experiment, the pixel size is 6.5 µm). The recovered phase is shown in Fig. 7d. Unlike our wavefront sensing technique, phase-shifting interferometry only provides modulo-$2\pi$ mappings (see the vertical scale in Fig. 6d), requiring the use of unwrapping algorithms to reconstruct a continuous phase like the one shown in Fig. 7b. We focus our attention on the small region between the two air bubbles in one of the photoresist stripes. After zooming in, it can be seen that the phase difference between the top of the strip and the substrate is 22.7 radians. In our measurements with the structured illumination setup, this difference is 23 radians, showing a very good agreement with the previous value. We also performed a thickness measurement of the region of interest with a mechanical profilometer (Dektak 6, Veeco). By using the strip height provided by the profilometer (3.8 µm), the refractive index of the photoresist layer can be estimated from the optical path length, $\left(L = \lambda \Delta \varphi / 2\pi \Delta n\right)$. In both phase shifting holography and our technique, the index estimation is 1.51, which is in good agreement with the nominal value of the photoresist material.

### D. Discussion

We have introduced a wavefront sensor based on patterned illumination produced by an amplitude spatial light modulator. Instead of resorting to an array of lenslets, as in SH wavefront sensing, the spatial information is captured by illuminating the object with binary amplitude masks generated with a DMD. The sensor is a simple position sensitive photodetector. This provides several benefits. First, we eliminate the spatial resolution constraints of traditional SH wavefront sensors. Second, by using the full size of the detector in each measurement, the dynamic range is greatly extended and the cross-talk problem is eliminated. Last, once the window size of the DMD has been fixed, changing the spatial resolution of the masks does not change the total amount of light, thus increasing optical efficiency when compared to SH wavefront sensing. This, in combination with the multiplex advantage, makes us think that the technique is well suited to work at low-light level scenarios, where similar approaches have been already proposed to obtain amplitude information [63]. Additionally, we have exploited the high spatial resolution offered by commercial DMDs to perform phase imaging with a pixel size of tens of microns. To this end, we have imaged a photoresist sample and calculated its local refraction index. These results have been demonstrated to be comparable with those provided by phase shifting interferometry, a well-established interferometric technique.

A distinctive feature of our technique is the fact that the spatial resolution is fixed by the spatial light modulator and the projecting optics, while the other optical elements included in the sensor, the focusing lens and the light detector, set the attainable sensitivity and dynamic range. By calculating several statistical moments of the light distribution at the detector plane (its irradiance and the position of its centroid), the method provides both amplitude and phase information. This enables the technique to produce results that are closer to the conventional phase imaging approaches, but still retaining some of the benefits of the SH approach. These benefits are the reference-less nature of SH wavefront sensing and the fact that the recovery process does not require phase unwrapping or iterative algorithms. In addition, the easy implementation of our system also offers the possibility of being used as an add-on module of a conventional microscope [20,64], thus allowing one to carry out quantitative phase imaging with sub-micrometric spatial resolutions.

Due to the temporal multiplexing nature of the technique, there is a tradeoff between spatial resolution and image acquisition time. Here, we have shown results with resolutions between $64 \times 64$ and $128 \times 128$ pixels, which take seconds to be acquired. Using faster, already available detectors, will allow image capturing at video frame rates. However, the greater the resolution, the slower the acquisition will be. This drawback can be considerably alleviated using CS [65] or adaptive approaches [66,67]. The feasibility of applying CS has been shown in Section 3, where a typical ocular aberration has been reconstructed with high compression ratios. In particular, acquisitions times of around 20 ms could be achieved operating at the full frame rate of the DMD, which would open up the door to perform real-time aberrometry with our system. Furthermore, this approach increases the global light efficiency of the method when compared to the raster scan procedure. Last, given the technological challenge of manufacturing both arrays of lenslets and pixelated sensors that work outside the visible spectrum, the technique proposed here is a good candidate to operate in regions such as IR and THz, where similar approaches have been already used to obtain amplitude images [52,68].

## 4. Methods

### A. Phase measurement from centroid position

Let us assume a light distribution, at position $z = 0$, with the form: $U_1(x,y) = A(x,y)e^{i\varphi(x,y)}$. If a thin lens, with a focal length $f$, is placed at the position $z = f$, the complex amplitude distribution at the position $z = 2f$, using Fresnel propagation, will be $U_2(x,y) = \frac{1}{j\lambda f}\widetilde{U}_1\left(\frac{x}{\lambda f}, \frac{y}{\lambda f}\right)$, where $\widetilde{U}_1$ denotes the Fourier transform of $U_1$. The irradiance of this light distribution will be given by $I_2 = |U_2(x,y)|^2 = \frac{1}{(\lambda f)^2}\left|\widetilde{U}_1\left(\frac{x}{\lambda f}, \frac{y}{\lambda f}\right)\right|^2$. Our technique is based on the calculus of several statistical moments of $I_2$. In particular, we will use the energy, $S$, and the centroid position, $\vec{\Delta} = (\Delta x; \Delta y)$. Those quantities will be given by:

$$S = \iint_{-\infty}^{\infty} I_2(x,y)\, dx\, dy, \quad (2)$$

$$\vec{\Delta} = \begin{pmatrix} \Delta x \\ \Delta y \end{pmatrix} = \frac{1}{S}\begin{pmatrix} \iint_{-\infty}^{\infty} x \cdot I_2(x,y)\, dx\, dy \\ \iint_{-\infty}^{\infty} y \cdot I_2(x,y)\, dx\, dy \end{pmatrix}. \quad (3)$$

By using Parseval's Theorem, it is easy to prove that $S = |U_2(x,y)|^2 = |U_1(x,y)|^2$, and the first measurement provides the energy of the wavefront. Using the Moment's Theorem of Fourier transformations and some algebra (see Supplement 1), it can be proved that the position of the centroid at the measurement plane and the phase of the object at the origin position are related by:

$$\vec{\Delta} = (\Delta x;\ \Delta y) = \frac{\lambda f}{2\pi S}\iint_{-\infty}^{\infty} A^2(x,y)\vec{\nabla}\varphi(x,y)\, dx\, dy. \quad (4)$$

For a small square aperture with lateral size $L$, placed at position $(a,b)$, the amplitude of the field can be described as:

$$A(x,y) = K \cdot rect\left(\frac{x-a}{L}\right) \cdot rect\left(\frac{y-b}{L}\right), \quad (5)$$

where $K$ is a constant. Introducing this into Eq. (4), we get:

$$\Delta x = \frac{\lambda f K^2}{2\pi S}\int_{y=b-L/2}^{y=b+L/2}\int_{x=a-L/2}^{x=a+L/2}\partial_x[\varphi(x,y)]\, dx\, dy;$$

$$\Delta y = \frac{\lambda f K^2}{2\pi S}\int_{y=b-L/2}^{y=b+L/2}\int_{x=a-L/2}^{x=a+L/2}\partial_y[\varphi(x,y)]\, dx\, dy. \quad (6)$$

If the square aperture is small enough, the gradients of the phase can be considered constant over the integration domain. Then, the centroid positions results:

$$\vec{\Delta} = (\Delta x; \Delta y) = \frac{K^2 \lambda f}{2\pi}\frac{L^2}{S}\vec{\nabla}_{a,b}\varphi(x,y), \quad (7)$$

where $\vec{\nabla}_{a,b} = \left(\partial/\partial_x \hat{i};\ \partial/\partial_y \hat{j}\right)\Big|_{x=a,y=b}$ represents the bidimensional gradient evaluated at point $x = a,\ y = b$. By combining Eq. (5) and Eq. (2), it is easy to prove that $S = K^2 L^2$, and thus the centroid of the distribution and the phase of the object are related by:

$$\vec{\Delta} = (\Delta x; \Delta y) = \frac{\lambda f}{2\pi}\vec{\nabla}_{a,b}\varphi(x,y). \quad (8)$$

It is clear that, for a small region of the wavefront, the centroid of the resulting distribution provides information about the gradient of the phase in that region. If the small square aperture is displaced, one can measure the correspondent centroid positions and then estimate the gradient map of the wavefront. After that, numerical integration provides the wavefront at the original position. If multiple square masks are used at the same time, it is also possible to relate the measurements with the phase at each one of the mask positions. Then, it is possible to understand Hadamard illumination as a particular case of this procedure, and the phase recovery can be performed while illuminating the full scene with a set of Hadamard patterns (see Supplement 1).

### B. Object recovery by multiplexing with Hadamard patterns

Every measurement can be mathematically described by the equation

$$\vec{y} = M\vec{x}, \quad (9)$$

where the object to be measured is described by the object vector $\vec{x}$, the measurements made are represented by the vector $\vec{y}$, and the measurement process is carried away by the sensing operator, represented by the matrix $M$. Once the measurements have been done, one just needs to solve the algebraic problem presented by Eq. (9) to recover the object. The structure of the sensing operator will be determined by both the nature of the object under study and the system used to recover an image. For a traditional camera setup, matrix $M$ is just the identity matrix, and each element of $\vec{x}$ is proportional to the energy of a given area of the object. By placing a detector at the image plane, the energies of each zone are measured and the object is recovered. Usually this is done with a sensor array, like a CCD camera, and all the measurements are done at the same time. However, there are scenarios, such as confocal microscopy, where this process is carried away by a bucket detector, usually a photomultiplier tube. In this case, the measurement process is sequential. It is possible to exploit Eq. (9) to increase the capabilities of an optical setup. In our approach, instead of just using the identity matrix as the sensing operator, each row of the matrix $M$ contains a Hadamard function. Now, the measurement process consists of making the superposition between the object and each one of the Hadamard functions. After that, the energy of that superposition is measured by the detector. This process can be made by projecting the functions onto the object with a spatial light modulator, as is done in our phase imaging system. When using this approach, each measurement contains light coming from several regions of the object. Doing this increases the amount of light at each measurement, and thus the signal-to-noise ratio. As Hadamard functions conform an orthonormal basis, Eq. (9) can be easily solved.

Once this way of measuring has been introduced, it is easy to see that it is possible to recover an image of different physical parameters, provided that the measurement process can be described by Eq. (9). It can be noted that our centroid measurements can be arranged in matrix form as:

$$\vec{\Delta} = \frac{\lambda f}{2\pi} M \overrightarrow{\vec{\nabla}\varphi}. \quad (10)$$

Here, $\vec{\Delta}$ is the vector containing the measured positions of the centroids, and now the object is the spatial gradient distribution, expressed as a vector, $\overrightarrow{\vec{\nabla}\varphi}$ (each element of this vector is a gradient for a given illumination pattern). Again, in our experiments we used the Hadamard functions as rows of the sensing matrix. For each Hadamard function projected onto the object, the energy of the distribution and the position of its centroid are measured with the position sensing detector. Then, Eq. (10) is solved and the gradient of the phase is obtained. After that, numerical integration provides the phase of the wavefront, $\varphi(x, y)$.

### C. Compressive Sensing

Solving the problem given by Eq. (9) is easy when sampling the object at the Shannon-Nyquist ratio. However, for $k < N$ measurements the equation system presents infinite solutions. The fundamental idea of CS is to use the fact that most of the signals found in nature are sparse in some basis of functions, i.e. they have a representation in some basis where most of the expansion coefficients are zero, so only a small number of them contain most of the relevant information. This fact can be used to provide a solution for the above underdetermined equation system. We can express the measurements as $\vec{y} = M\vec{x} = M\Psi\vec{\alpha} = \Phi\vec{\alpha}$, where we have represented the object in some basis of functions ($\vec{x} = \Psi\vec{\alpha}$). Then, CS algorithms provide a solution to the $l_1$-norm minimization problem:

$$\vec{\alpha} = arg\, min \|\overrightarrow{\alpha'}\|_1 \text{ such that } \Phi\overrightarrow{\alpha'} = \vec{y}. \quad (11)$$

In order to find a good solution, some premises need to be fulfilled [69]. First, the number of measurements cannot be arbitrarily low. Depending on the sparsity of the object (which depends on the chosen recovery basis), higher or lower number of measurements are needed to obtain good results. Second, the basis pair (measurement and recovery) need to be incoherent. In practice, it is usually easy to find pairs of basis that fulfill this principle. Usually, the measurement basis is chosen for its ease of generation in a SLM, and once one has been picked, the user selects a recovery basis that fulfills the incoherence constraint and where the object is sparse. Last, a sensing strategy needs to be defined. Initially, it was proposed that randomly choosing a subset of elements in the measurement basis was enough to obtain good results [43]. Soon after that, other strategies were defined to reduce the number of measurements maintaining the image quality. For example, it is known that natural scenes tend to have a power spectrum centered around low frequencies, so mixed sampling strategies have been used with very good results [58].

In our experiments, $M$ represents the shifted and rescaled Hadamard basis (which elements are either 0 or 1), easily generated on a DMD, and $\Psi$ is chosen to be the Haar wavelet basis, where most images tend to be sparse. The measurements have been performed following the mixed approach: fixed the number of samples, $k$, we chose the lowest frequency $k/2$ Hadamard patterns, and then we select $k/2$ random Hadamard patterns from the remaining elements of the complete basis.

### D. Phase shifting holography measurements

For the measurements using phase shifting digital holography, we used an interferometer in a Mach-Zehnder configuration. A collimated laser (Oxxius slim-532) was divided by a beam splitter into the object and the reference beam. The first one illuminated the object while the second traveled directly towards the camera (Allied Stingray F-145). The object, a layer of photoresist spin coated onto a transparent plate, was imaged onto the camera by an optical system in a 4-f configuration. The phase shifts were generated by shifting a grating codified onto a DMD (DLP Discovery 4100) and filtering the first diffraction order in the Fourier plane of a 4-f optical configuration located before the object plane [70]. The camera recorded four interferograms with a phase shift interval equal to π/2 and a standard phase-shifting algebraic operation [71] was used to measure the amplitude and phase at the object plane.

**Funding**. MINECO (FIS2016-75618-R and FIS2015-72872-EXP), Generalitat Valenciana (PROMETEO/2016/079), and Universitat Jaume I (P1-1B2015-35, PREDOC/2013/32).

**Acknowledgment**. We thank Salvador Bará for the photoresist plates simulating optical aberrations used in our experiments.

**Author contributions.** F.S. and E.T. conceived the idea; F.S. performed the experiments with the aid of P.C. and V. D.; F.S. and V.D. wrote the manuscript; all the authors discussed the results and revised the manuscript; J.L. and E.T. coordinated the project.

See Supplement 1 for supporting content.

**REFERENCES**

1. F. Zernike, "How I Discovered Phase Contrast," Science (80-. ). **121**, 345–349 (1955).
2. G. Popescu, "The power of imaging with phase, not power," Phys. Today **70**, 34–40 (2017).
3. J. Liang, B. Grimm, S. Goelz, and J. F. Bille, "Objective measurement of wave aberrations of the human eye with the use of a Hartmann–Shack wave-front sensor," J. Opt. Soc. Am. A **11**, 1949–1957 (1994).
4. P. Artal, "Optics of the eye and its impact in vision : a tutorial," Adv. Opt. Photonics **6**, 340–367 (2014).
5. G. Nehmetallah and P. P. Banerjee, "Applications of digital and analog holography in three-dimensional imaging," Adv. Opt. Photonics **4**, 472–553 (2012).
6. D. Gabor, "Microscopy by Reconstructed Wave-Fronts," Proc. R. Soc. A Math. Phys. Eng. Sci. **197**, 454–487 (1949).
7. M. Mir, B. Bhaduri, R. Wang, R. Zhu, and G. Popescu, "Quantitative Phase Imaging," in *Progress in Optics* (Elsevier Inc., 2012), Vol. 57, pp. 133–217.
8. R. Tyson, *Principles of Adaptive Optics, Third Edition*, Series in Optics and Optoelectronics (CRC Press, 2010), Vol. 20102628.
9. M. J. Booth, "Adaptive optical microscopy: the ongoing quest for a perfect image," Light Sci. Appl. **3**, e165 (2014).
10. N. Ji, "Adaptive optical fluorescence microscopy," Nat. Methods **14**, 374–380 (2017).
11. K. Wang, D. E. Milkie, A. Saxena, P. Engerer, T. Misgeld, M. E. Bronner, J. Mumm, and E. Betzig, "Rapid adaptive optical recovery of optimal resolution over large volumes," Nat. Methods **11**, 625–628 (2014).
12. A. Kumar, W. Drexler, and R. A. Leitgeb, "Subaperture correlation based digital adaptive optics for full field optical coherence tomography," Opt. Express **21**, 10850–10866 (2013).
13. Y. Shechtman, Y. C. Eldar, O. Cohen, H. N. Chapman, J. Miao, and M. Segev, "Phase Retrieval with Application to Optical Imaging: A contemporary overview," IEEE Signal Process. Mag. **32**, 87–109 (2015).
14. G. Zheng, R. Horstmeyer, and C. Yang, "Wide-field, high-resolution Fourier ptychographic microscopy," Nat. Photonics **7**, 739–745 (2013).
15. H. N. Chapman and K. A. Nugent, "Coherent lensless X-ray imaging," Nat. Photonics **4**, 833–839 (2010).
16. R. Horisaki, Y. Ogura, M. Aino, and J. Tanida, "Single-shot phase imaging with a coded aperture," Opt. Lett. **39**, 6466 (2014).
17. N. Streibl, "Phase imaging by the transport equation of intensity," Opt. Commun. **49**, 6–10 (1984).
18. R. Horisaki, R. Egami, and J. Tanida, "Single-shot phase imaging with randomized light (SPIRaL)," Opt. Express **24**, 3765 (2016).
19. L. Tian, X. Li, K. Ramchandran, and L. Waller, "Multiplexed coded illumination for Fourier Ptychography with an LED array

20. L. Tian and L. Waller, "3D intensity and phase imaging from light field measurements in an LED array microscope," Optica **2**, 104 (2015).
21. J. J. Field, K. A. Wernsing, S. R. Domingue, A. M. Allende Motz, K. F. DeLuca, D. H. Levi, J. G. DeLuca, M. D. Young, J. A. Squier, and R. A. Bartels, "Superresolved multiphoton microscopy with spatial frequency-modulated imaging," Proc. Natl. Acad. Sci. **113**, 6605–6610 (2016).
22. P. Gao, G. Pedrini, and W. Osten, "Phase retrieval with resolution enhancement by using structured illumination," Opt. Lett. **38**, 5204–5207 (2013).
23. B. C. Platt and R. Shack, "History and principles of Shack-Hartmann wavefront sensing.," J. Refract. Surg. **17**, S573–S577 (2001).
24. B. Stoklasa, L. Motka, J. Rehacek, Z. Hradil, and L. L. Sánchez-Soto, "Wavefront sensing reveals optical coherence," Nat. Commun. **5**, 3275 (2014).
25. R. W. Wilson, "SLODAR: measuring optical turbulence altitude with a Shack-Hartmann wavefront sensor," Mon. Not. R. Astron. Soc. **337**, 103–108 (2002).
26. D. Dayton, J. Gonglewski, B. Pierson, and B. Spielbusch, "Atmospheric structure function measurements with a Shack–Hartmann wave-front sensor," Opt. Lett. **17**, 1737–1739 (1992).
27. X. Cui, J. Ren, G. J. Tearney, and C. Yang, "Wavefront image sensor chip," Opt. Express **18**, 16685 (2010).
28. L. Seifert, J. Liesener, and H. J. Tiziani, "The adaptive Shack–Hartmann sensor," Opt. Commun. **216**, 313–319 (2003).
29. Y. Saita, H. Shinto, and T. Nomura, "Holographic Shack–Hartmann wavefront sensor based on the correlation peak displacement detection method for wavefront sensing with large dynamic range," Optica **2**, 411–415 (2015).
30. S. Thomas, T. Fusco, A. Tokovinin, M. Nicolle, V. Michau, and G. Rousset, "Comparison of centroid computation algorithms in a Shack-Hartmann sensor," Mon. Not. R. Astron. Soc. **371**, 323–336 (2006).
31. J. Pfund, N. Lindlein, and J. Schwider, "Dynamic range expansion of a Shack–Hartmann sensor by use of a modified unwrapping algorithm," Opt. Lett. **23**, 995–997 (1998).
32. S. Wang, P. Yang, B. Xu, L. Dong, and M. Ao, "Shack-Hartmann wavefront sensing based on binary-aberration-mode filtering," Opt. Express **23**, 5052–5063 (2015).
33. R. Martínez-Cuenca, V. Durán, V. Climent, E. Tajahuerce, S. Bará, J. Ares, J. Arines, M. Martínez-Corral, and J. Lancis, "Reconfigurable Shack-Hartmann sensor without moving elements.," Opt. Lett. **35**, 1338–1340 (2010).
34. T. Godin, M. Fromager, E. Cagniot, M. Brunel, and K. Aït-Ameur, "Reconstruction-free sensitive wavefront sensor based on continuous position sensitive detectors," Appl. Opt. **52**, 8310–8317 (2013).
35. R. Navarro and E. Moreno-Barriuso, "Laser ray-tracing method for optical testing," Opt. Lett. **24**, 951 (1999).
36. S. Olivier, V. Laude, and J. Huignard, "Liquid-crystal Hartmann wave-front scanner," Appl. Opt. **39**, 3838 (2000).
37. J.-C. Chanteloup, "Multiple-wave lateral shearing interferometry for wave-front sensing," Appl. Opt. **44**, 1559 (2005).
38. "PHASICS," http://phasicscorp.com.
39. I. Iglesias, "Pyramid phase microscopy," Opt. Lett. **36**, 3636 (2011).
40. A. B. Parthasarathy, K. K. Chu, T. N. Ford, and J. Mertz, "Quantitative phase imaging using a partitioned detection aperture," Opt. Lett. **37**, 4062 (2012).
41. I. Iglesias and F. Vargas-Martin, "Quantitative phase microscopy of transparent samples using a liquid crystal display," J. Biomed. Opt. **18**, 26015 (2013).
42. H. Lu, J. Chung, X. Ou, and C. Yang, "Quantitative phase imaging and complex field reconstruction by pupil modulation differential phase contrast," Opt. Express **24**, 25345 (2016).
43. E. J. J. Candes and M. B. Wakin, "An Introduction To Compressive Sampling," IEEE Signal Process. Mag. **25**, 21–30 (2008).
44. M. F. F. Duarte, M. A. A. Davenport, D. Takhar, J. N. N. Laska, K. F. F. Kelly, and R. G. Baraniuk, "Single-Pixel Imaging via Compressive Sampling," IEEE Signal Process. Mag. **25**, 83–91 (2008).
45. H. Gong, O. Soloviev, D. Wilding, P. Pozzi, M. Verhaegen, and G. Vdovin, "Holographic imaging with a Shack-Hartmann wavefront sensor," Opt. Express **24**, 13729–13737 (2016).
46. F. Zhang, G. Pedrini, and W. Osten, "Phase retrieval of arbitrary complex-valued fields through aperture-plane modulation," Phys. Rev. A **75**, 43805 (2007).
47. P. Gao, G. Pedrini, C. Zuo, and W. Osten, "Phase retrieval using spatially modulated illumination," Opt. Lett. **39**, 3615 (2014).
48. G. Vdovin, H. Gong, O. Soloviev, P. Pozzi, and M. Verhaegen, "Lensless coherent imaging by sampling of the optical field with digital micromirror device," J. Opt. **17**, 122001 (2015).
49. W. H. Southwell, "Wave-front estimation from wave-front slope measurements," J. Opt. Soc. Am. **70**, 998–1006 (1980).
50. R. Kasztelanic, A. Filipkowski, D. Pysz, R. Stepien, A. J. Waddie, M. R. Taghizadeh, and R. Buczynski, "High resolution Shack-Hartmann sensor based on array of nanostructured GRIN lenses," Opt. Express **25**, 1680–1691 (2017).
51. M. Saxena, G. Eluru, and S. S. Gorthi, "Structured illumination microscopy," Adv. Opt. Photonics **7**, 241–275 (2015).
52. C. M. Watts, D. Shrekenhamer, J. Montoya, G. Lipworth, J. Hunt, T. Sleasman, S. Krishna, D. R. Smith, and W. J. Padilla, "Terahertz compressive imaging with metamaterial spatial light modulators," Nat. Photonics **8**, 605–609 (2014).
53. F. Soldevila, E. Irles, V. Durán, P. Clemente, M. Fernández-Alonso, E. Tajahuerce, and J. Lancis, "Single-pixel polarimetric imaging spectrometer by compressive sensing," Appl. Phys. B **113**, 551–558 (2013).
54. F. Soldevila, P. Clemente, E. Tajahuerce, N. Uribe-Patarroyo, P. Andrés, and J. Lancis, "Computational imaging with a balanced detector," Sci. Rep. **6**, 29181 (2016).
55. E. Tajahuerce, V. Durán, P. Clemente, E. Irles, F. Soldevila, P. Andrés, and J. Lancis, "Image transmission through dynamic scattering media by single-pixel photodetection," Opt. Express **22**, 16945–16955 (2014).
56. V. Durán, F. Soldevila, E. Irles, P. Clemente, E. Tajahuerce, P. Andrés, and J. Lancis, "Compressive imaging in scattering media," Opt. Express **23**, 14424–14433 (2015).
57. P. R. Griffiths, H. J. Sloane, and R. W. Hannah, "Interferometers vs Monochromators: Separating the Optical and Digital Advantages," Appl. Spectrosc. **31**, 485–495 (1977).
58. V. Studer, J. Bobin, M. Chahid, H. S. Mousavi, E. Candes, and M. Dahan, "Compressive fluorescence microscopy for biological and hyperspectral imaging.," Proc. Natl. Acad. Sci. U. S. A. **109**, E1679–E1687 (2012).
59. M. Ducros, Y. Goulam Houssen, J. Bradley, V. de Sars, and S. Charpak, "Encoded multisite two-photon microscopy.," Proc. Natl. Acad. Sci. U. S. A. **110**, 13138–13143 (2013).
60. W. Pratt, J. Kane, and H. Andrews, "Hadamard transform image coding," Proc. IEEE **57**, 58–67 (1969).
61. E. Candès and J. Romberg, "l1-magic : Recovery of Sparse Signals via Convex Programming," http://statweb.stanford.edu/~candes/l1magic/downloads/l1magic.pdf.
62. H. Gong, T. E. Agbana, P. Pozzi, O. Soloviev, M. Verhaegen, and G. Vdovin, "Optical path difference microscopy with a Shack–Hartmann wavefront sensor," Opt. Lett. **42**, 2122–2125 (2017).
63. P. A. Morris, R. S. Aspden, J. E. C. Bell, R. W. Boyd, and M. J. Padgett, "Imaging with a small number of photons," Nat. Commun. **6**, 5913 (2015).
64. Z. Wang, L. Millet, M. Mir, H. Ding, S. Unarunotai, J. Rogers, M. U. Gillette, and G. Popescu, "Spatial light interference microscopy (SLIM)," Opt. Express **19**, 1016–1026 (2011).
65. E. Candès, "Compressive sampling," in *Proceedings of the International Congress of Mathematicians Madrid, August 22–30, 2006* (European Mathematical Society Publishing House, 2006), pp. 1433–1452.
66. F. Soldevila, E. Salvador-Balaguer, P. Clemente, E. Tajahuerce, and J. Lancis, "High-resolution adaptive imaging with a single photodiode," Sci. Rep. **5**, 14300 (2015).
67. H. Dai, G. Gu, W. He, L. Ye, T. Mao, and Q. Chen, "Adaptive compressed photon counting 3D imaging based on wavelet trees and depth map sparse representation," Opt. Express **24**, 26080–26096 (2016).

68. N. Radwell, K. J. Mitchell, G. GIbson, M. Edgar, R. Bowman, and M. J. Padgett, "Single-pixel infrared and visible microscope," Optica **1**, 285–289 (2014).
69. J. Romberg, "Imaging via Compressive Sampling," IEEE Signal Process. Mag. **25**, 14–20 (2008).
70. D. B. Conkey, A. M. Caravaca-Aguirre, and R. Piestun, "High-speed scattering medium characterization with application to focusing light through turbid media.," Opt. Express **20**, 1733–40 (2012).
71. I. Yamaguchi and T. Zhang, "Phase-shifting digital holography," Opt. Lett. **22**, 1268–1270 (1997).

# Phase imaging by spatial wavefront sampling: supplementary material

F. SOLDEVILA,[1,*] V. DURÁN,[2,3] P. CLEMENTE,[1,4] J. LANCIS,[1] E. TAJAHUERCE[1]

[1]GROC·UJI, Institute of New Imaging Technologies (INIT), Universitat Jaume I, E12071 Castelló, Spain.
[2] Univ. Grenoble Alpes, LIPHY, F-38000 Grenoble, France.
[3]CNRS, LIPHY, F-38000 Grenoble, France.
[4]Servei Central d'Instrumentació Científica (SCIC), Universitat Jaume I, E12071 Castelló, Spain.
*Corresponding author: fsoldevi@uji.es

This document provides supplementary information to "Phase imaging by spatial wavefront sampling,". Figures including additional experimental results and extended mathematical formulation related to the phase measurement method used in the paper are included.

## Measurement of an optical aberration and comparison with Shack-Hartman wavefront sensing and phase shifting interferometry

Fig. S1 shows the results for the plate that generates a coma aberration. Using our approach (experimental setup of Fig. 3a in the manuscript), the patterns projected onto the plate had a resolution of 64×64 pixels. As a way to smooth the retrieved phase distribution, the Zernike decomposition of the wavefront was calculated. Then, an image with high spatial resolution (512×512) was generated with that information. This process made it possible to compare that phase image with the one obtained with the Shack-Hartmann sensor used in our experiments, since the commercial software of this sensor directly calculates the Zernike decomposition and also provides a smoothed high-resolution image. Phase shifting interferometry results were obtained by applying four different phase shifts between the two arms of a Mach-Zehnder interferometer (see methods). The phase distributions provided by the three methods present high resemblance, as can be observed in Fig. S1. For our approach, the peak-to-valley phase value is 6.96 λ. The Shack-Hartmann sensor provides a value of 7.24 λ, while the corresponding value for phase measured by phase-shifting interferometry is 7.20 λ. The discrepancy between the above quantities is less than 5% of the total peak-to-valley phase value obtained with phase shifting interferometry, confirming the validity of the approach.

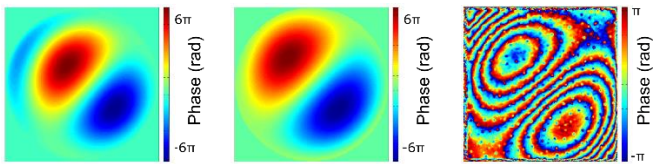

Fig. S1 Phase introduced by a plate simulating a coma aberration. Phase retrieved using wavefront spatial modulation, as shown in Fig. 3 of the manuscript (left). Same phase distribution obtained with a Shack-Hartmann sensor (middle). Wrapped phase measured via phase-shifting interferometry (right).

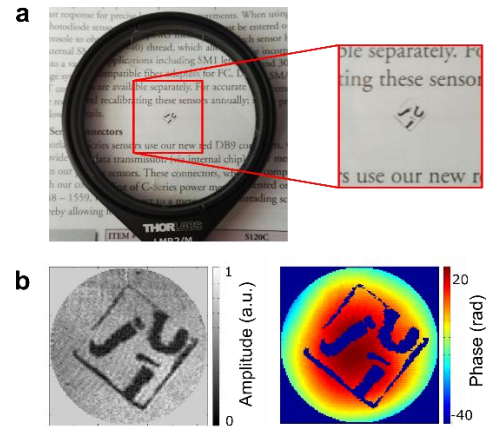

Fig. S2 Amplitude and phase recovery. (a) Image of the object under study, a positive lens whose central part is covered by an amplitude mask. (b) Spatial distribution of the reconstructed complex amplitude. Given that the mask has opaque regions where the phase is not defined, we arbitrarily assign a null phase value to those object zones.

## Complex amplitude retrieval of an object

Here we show a simple example of the retrieval of an amplitude and phase object with our technique. We used an object composed of an amplitude mask attached to the lens used in section 3 of the manuscript (Fig. S2a). The reconstruction basis is formed by Hadamard functions of $128 \times 128$ pixels, with a pixel pitch of $41.04$ µm. The retrieved complex amplitude is shown in Fig. S2. As explained in the Methods section of the manuscript, after the position detector provides the total irradiance and the centroid position of the resulting light distribution for the full set of Hadamard patterns projected onto the sample. Using the irradiances provides the amplitude image (left inset of Fig. S2b), and the position of the centroids can be used to obtain the phase information (right inset of Fig. S2b).

# Relation between the phase of an object and the high-order statistical moments of its intensity Fourier distribution

Here we will discuss the relation between the phase of a wavefront in a given plane and the high-order statistical moments of the intensity distribution at the Fourier plane of a lens. This configuration is represented in Fig. S3. In order to work out the relation, Parseval's Theorem and the Moment's Theorem of Fourier Transforms will be invoked.

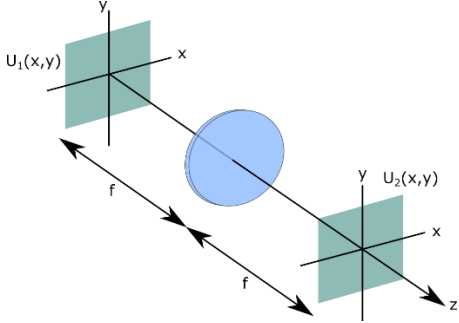

Fig. S3. Optical configuration of the derivation of Eqs. (5) and (22). The field distribution at the first plane is related to the field distribution at the second plane through a thin lens. Both planes are separated a distance twice the focal length, f, of the lens.

We assume the initial field to have the following form: $U_1(x,y) = A(x,y)e^{j\varphi(x,y)}$. By using Fresnel's propagation, we can link the field in the first plane, $U_1(x,y)$, with the field in the second plane, $U_2(x,y)$, in the following way:

$$U_2(x,y) = \frac{1}{j\lambda f}\tilde{U}_1\left(\frac{x}{\lambda f},\frac{y}{\lambda f}\right), \quad (1)$$

where $\tilde{U}_1$ denotes the Fourier transform of $U_1$. As the planes we are working with are both at a distance $f$ from the lens, there is no global phase involved.

Given that we measure the light irradiance, the relevant quantity is the square modulus of this distribution:

$$I_2 = |U_2(x,y)|^2 = \frac{1}{(\lambda f)^2}\left|\tilde{U}_1\left(\frac{x}{\lambda f},\frac{y}{\lambda f}\right)\right|^2. \quad (2)$$

Our detector will provide the total energy of this distribution, which we will call $S$, and the centroid of the distribution, which we will call $\vec{\Delta} = (\Delta x; \Delta y)$. Those physical measurements can be mathematically described with the following operations:

$$S = \iint_{-\infty}^{\infty} I_2(x,y)\,dx\,dy; \quad (3)$$

$$\vec{\Delta} = \begin{pmatrix}\Delta x\\ \Delta y\end{pmatrix} = \frac{1}{S}\begin{pmatrix}\iint_{-\infty}^{\infty} x\cdot I_2(x,y)\,dx\,dy\\ \iint_{-\infty}^{\infty} y\cdot I_2(x,y)\,dx\,dy\end{pmatrix}. \quad (4)$$

Using Parseval's Theorem:

$$S = \iint_{-\infty}^{\infty}\frac{1}{(\lambda f)^2}\left|\tilde{U}_1\left(\frac{x}{\lambda f},\frac{y}{\lambda f}\right)\right|^2 dx\,dy = \iint_{-\infty}^{\infty}|U_1(x,y)|^2 dx\,dy, \quad (5)$$

that is, $S$ is just the total energy at the first plane. Now, let us make the calculation of the centroid of the distribution. By using its definition, we have, for $\Delta x$:

$$\Delta x = \frac{1}{S}\iint_{-\infty}^{\infty} x\cdot I_2(x,y)\,dx\,dy =$$
$$= \frac{1}{S}\iint_{-\infty}^{\infty} x\frac{1}{(\lambda f)^2}\left|\tilde{U}_1\left(\frac{x}{\lambda f},\frac{y}{\lambda f}\right)\right|^2 dx\,dy. \quad (6)$$

After a variable change ($x' = x/\lambda f$, $y' = y/\lambda f$) we obtain:

$$\Delta x = \frac{\lambda f}{S}\iint_{-\infty}^{\infty} x'|\tilde{U}_1(x',y')|^2 dx'\,dy'. \quad (7)$$

For the sake of simplicity, from now on the marks in the spatial variables inside the integral will be removed. In order to solve the integral, we will invoke the Moment's Theorem of Fourier transformations and some properties of Fourier transformations. First, the Moment's Theorem states [1]:

$$m_{k,l} = \iint_{-\infty}^{\infty} u^k v^l F(u,v)\,du\,dv = \frac{f^{(k,l)}(x,y)}{(j2\pi)^{(k+l)}}\Bigg|_{\alpha=\beta=0}$$
$$= \frac{f^{(k,l)}(0,0)}{(j2\pi)^{(k+l)}}; \quad (8)$$

where

$$f^{(k,l)}(x,y) = \frac{\partial^l\partial^k f(x,y)}{\partial y^l \partial x^k}, \quad (9)$$

and $f$ and $F$ are related via Fourier transform.

Now, we see that Eq. (7) involves the first moment, $m_{1,0}$ ($k=1, l=0$). In order to solve the integral, we will also need the Fourier transform of $|\tilde{U}_1(x,y)|^2$. It is easy to prove that [1]:

$$\mathcal{F}\left\{|\tilde{U}_1(x,y)|^2\right\} = U_1(u,v) \star\star U_1^*(u,v). \quad (10)$$

In other words, the Fourier transform of the square modulus of a function is the complex cross correlation of its Fourier transform. Combining Eq. (7) and Eq. (8), we get:

$$\Delta x = \frac{\lambda f}{S}\iint_{-\infty}^{\infty} x|\tilde{U}_1(x,y)|^2 dx\,dy = \frac{\lambda f}{S}\frac{1}{j2\pi}f^{(1,0)}(0,0)$$
$$= \frac{\lambda f}{j2\pi S}\frac{\partial}{\partial u}\left[\mathcal{F}\left\{|\tilde{U}_1(x,y)|^2\right\}\right]\Bigg|_{u=v=0}; \quad (11)$$

Taking into account Eq. (10):

$$\Delta x = \frac{\lambda f}{j2\pi S}\frac{\partial}{\partial u}\left[\iint_{-\infty}^{\infty} U_1(p+u,q+v)U_1^*(p,q)\,dp\,dq\right]\Bigg|_{u=v=0}. \quad (12)$$

Introducing the derivative into the integral:

$$\Delta x = \frac{\lambda f}{j2\pi S}\iint_{-\infty}^{\infty}\frac{\partial}{\partial u}[U_1(p+u,q+v)]\Bigg|_{u=v=0}U_1^*(p,q)\,dp\,dq. \quad (13)$$

The derivative will be, after the variable change $x = p+u, dx = du$:

$$\frac{\partial}{\partial u}[U_1(p+u,q+v)]\Bigg|_{u=v=0}$$
$$= \partial_u U_1(p+u,q+v)|_{u=v=0} = \quad (14)$$
$$= \frac{\partial}{\partial x}[U_1(x,q+v)]\Bigg|_{x=p,v=0} = \partial_x U_1(p,q).$$

So, what we get is the partial derivative of the complex field in one direction. In this way, $\Delta x$ can be expressed as:

$$\Delta x = \frac{\lambda f}{j2\pi S}\iint_{-\infty}^{\infty}[\partial_x U_1(x,y)]\,U_1^*(x,y)dx\,dy. \quad (15)$$

In a similar way, $\Delta y$ can be expressed as:

$$\Delta y = \frac{\lambda f}{j2\pi S}\iint_{-\infty}^{\infty}[\partial_y U_1(x,y)]\,U_1^*(x,y)dx\,dy. \quad (16)$$

Introducing the analytic form of $U_1$:

$$U_1(x,y) = A(x,y)e^{j\varphi(x,y)};$$
$$U_1^*(x,y) = A(x,y)e^{-j\varphi(x,y)}; \quad (17)$$

$$\partial_x U_1(x,y) = [\partial_x A(x,y)]e^{j\varphi(x,y)} + j[\partial_x \varphi(x,y)]A(x,y)e^{j\varphi(x,y)}.$$

Then, $\Delta x$ results:

$$\Delta x = \frac{\lambda f}{j 2\pi S}\left[\iint_{-\infty}^{\infty} [\partial_x A(x,y)]e^{j\varphi(x,y)}A(x,y)e^{-j\varphi(x,y)}\,dx\,dy + j\iint_{-\infty}^{\infty} A^2(x,y)[\partial_x \varphi(x,y)]\,dx\,dy\right]. \quad (18)$$

The first term is zero for two main reasons. First, for the centroid position to be real. Second, the integrand is a perfect differential that integrates to zero, given that the amplitude goes to zero at infinity. It can easily be seen that:

$$[\partial_x A(x,y)]A(x,y)\,dx = \partial_x\left[\frac{1}{2}A^2(x,y)\right]dx. \quad (19)$$

Then, for the first integral, one has:

$$\int_{-\infty}^{\infty} \partial_x\left[\frac{1}{2}A^2(x,y)\right]dx = \frac{1}{2}A^2(x,y)\Big|_{x=-\infty}^{x=\infty} = 0. \quad (20)$$

And $\Delta x$ will be:

$$\Delta x = \frac{\lambda f}{2\pi S}\iint_{-\infty}^{\infty} A^2(x,y)\,[\partial_x \varphi(x,y)]\,dx\,dy. \quad (21)$$

The same procedure can be used to calculate $\Delta y$, and introducing the gradient operator, $\vec{\nabla} = (\Delta x; \Delta y) = \left(\hat{\imath}\,\partial/\partial x;\hat{\jmath}\,\partial/\partial y\right)$, the centroid position can be described as:

$$\vec{\Delta} = (\Delta x; \Delta y) = \frac{\lambda f}{2\pi S}\iint_{-\infty}^{\infty} A^2(x,y)\,\vec{\nabla}[\varphi(x,y)]\,dx\,dy. \quad (22)$$

So, the centroid position of the intensity distribution in the detector plane is related to the gradient of the phase of the wavefront in the initial plane. By using Eq. (22), it is possible to relate different centroid measurements to the phase of the wavefront, as we will see in the next section.

## Using a square mask to measure the local phase of the wavefront

Let us describe what happens if we sample the complex field with an amplitude square mask that only allows the light to pass through a square window of size $L$ at a position $(a, b)$. The configuration is shown in Fig. S4.

Then, the field can be described by using the rectangular functions in the following way:

$$A(x,y) = K\,rect\left(\frac{x-a}{L}\right)rect\left(\frac{y-b}{L}\right). \quad (23)$$

Introducing this amplitude into the centroid expression, we get, for $\Delta x$:

$$\Delta x = \frac{\lambda f K^2}{2\pi S}\iint_{-\infty}^{\infty} rect\left(\frac{x-a}{L}\right)rect\left(\frac{y-b}{L}\right)[\partial_x \varphi(x,y)]\,dx\,dy. \quad (24)$$

Now, we can consider that if $L$ is small enough, the gradient of the phase of the object will be constant over the whole integration domain (i.e., the phase will be locally flat on that region). By doing this, the partial differentiation term goes out of the integral and we have:

$$\Delta x = \frac{\lambda f K^2}{2\pi S}\partial_x\varphi(x,y)|_{(a,b)}\int_{y=b-L/2}^{y=b+L/2}\int_{x=a-L/2}^{x=a+L/2} dx\,dy =$$

$$= \frac{\lambda f K^2}{2\pi S}\partial_x\varphi(x,y)|_{(a,b)}\,x\big|_{a-L/2}^{a+L/2}\,y\big|_{b-L/2}^{b+L/2} \quad (25)$$

$$= \frac{\lambda f}{2\pi}\frac{K^2 L^2}{S}\partial_x\varphi(a,b).$$

Doing the same for $\Delta y$, we obtain:

$$\Delta y = \frac{\lambda f}{2\pi}\frac{K^2 L^2}{S}\partial_y\varphi(a,b). \quad (26)$$

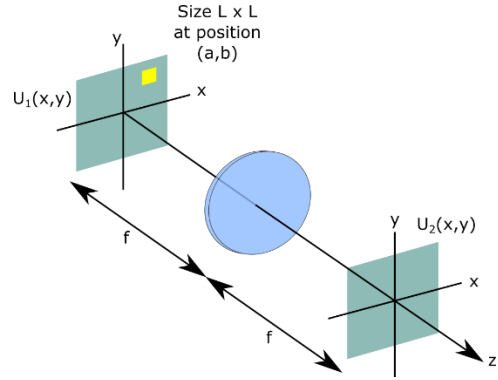

Fig. S4. Masking the wavefront with a square aperture. Now, only a small portion of the wavefront (the yellow part) is transmitted through the system.

In this case, it is easy to prove that $S$, the energy of the wavefront in the first plane, will be equal to $K^2 L^2$:

$$S = \iint_{-\infty}^{\infty} |U_1(x,y)|^2\,dx\,dy =$$

$$= K^2\iint_{-\infty}^{\infty} rect^2\left(\frac{x-a}{L}\right)rect^2\left(\frac{y-b}{L}\right)dx\,dy = K^2 L^2, \quad (27)$$

then, by introducing the operator $\vec{\nabla}_{(a,b)} = \left(\hat{\imath}\frac{\partial}{\partial x};\hat{\jmath}\frac{\partial}{\partial y}\right)\Big|_{(a,b)}$, we arrive to:

$$\vec{\Delta} = (\Delta x; \Delta y) = \frac{\lambda f}{2\pi}\vec{\nabla}_{(a,b)}\varphi(x,y). \quad (28)$$

So, for a square aperture placed in the position $(a, b)$, the value of the measured centroid will be proportional to the gradient of the object phase within the area of the aperture.

## Spatial multiplexing of the phase information of a wavefront

Now we will see what happens if we have more than one aperture at the same time. In this case, the problem will be as depicted in Fig. S5. Now, the field will be given by:

$$A(x,y) = K_1 rect\left(\frac{x-a}{L}\right)rect\left(\frac{y-b}{L}\right) + K_2 rect\left(\frac{x-c}{L}\right)rect\left(\frac{y-d}{L}\right). \quad (29)$$

The energy at the object plane will be:

$$S = \iint_{-\infty}^{\infty} |U_1(x,y)|^2\,dx\,dy = \iint_{-\infty}^{\infty} |A(x,y)|^2\,dx\,dy. \quad (30)$$

And,

$$|A(x,y)|^2 =$$
$$= K_1^2 rect^2\left(\frac{x-a}{L}\right)rect^2\left(\frac{y-b}{L}\right)$$
$$+ K_2^2 rect^2\left(\frac{x-c}{L}\right)rect^2\left(\frac{y-d}{L}\right) \quad (31)$$
$$+ 2K_1 K_2 rect\left(\frac{x-a}{L}\right)rect\left(\frac{y-b}{L}\right)rect\left(\frac{x-c}{L}\right)rect\left(\frac{y-d}{L}\right).$$

The crossed term will be always zero if both masks do not overlap. This is the case we will always consider: several square masks acting onto the field at different positions. Then the energy will be:

$$S = K_1^2\iint_{-\infty}^{\infty} rect^2\left(\frac{x-a}{L}\right)rect^2\left(\frac{y-b}{L}\right)dx\,dy +$$
$$+ K_2^2\iint_{-\infty}^{\infty} rect^2\left(\frac{x-c}{L}\right)rect^2\left(\frac{y-d}{L}\right)dx\,dy \quad (32)$$
$$= L^2(K_1^2 + K_2^2).$$

which is the sum of the energy of the two individual masks, as one can expect. Now, the centroid will be:

$$\Delta x = \frac{\lambda f}{2\pi S} \iint_{-\infty}^{\infty} A^2(x,y) [\partial_x \varphi(x,y)] dx\, dy =$$

$$= \frac{\lambda f}{2\pi S} \left[ K_1^2 \iint_{-\infty}^{\infty} rect^2\left(\frac{x-a}{L}\right) rect^2\left(\frac{y-b}{L}\right) \partial_x \varphi(x,y)\, dx \right. \quad (33)$$

$$\left. + K_2^2 \iint_{-\infty}^{\infty} rect^2\left(\frac{x-c}{L}\right) rect^2\left(\frac{y-d}{L}\right) \partial_x \varphi(x,y)\, dx\, dy \right].$$

We can see that this this result is similar to that obtained for the single-aperture one, but now with two integrals. By solving both integrals, $\Delta x$ can be expressed as:

$$\Delta x = \frac{\lambda f}{2\pi} \left[ \frac{L^2 K_1^2}{L^2(K_1^2 + K_2^2)} \partial_x \varphi(x,y)|_{x=a, y=b} \right. \quad (34)$$
$$\left. + \frac{L^2 K_2^2}{L^2(K_1^2 + K_2^2)} \partial_x \varphi(x,y)|_{x=c, y=d} \right].$$

If we identify $L^2 K_i = M_i$ as the energy of each square aperture, and $L^2(K_1^2 + K_1^2)$ as the energy of all the apertures combined, we have:

$$\Delta x = \frac{\lambda f}{2\pi} \left[ \frac{M_1}{M_1 + M_2} \partial_x \varphi(x,y)|_{x=a,y=b} \right. \quad (35)$$
$$\left. + \frac{M_2}{M_1 + M_2} \partial_x \varphi(x,y)|_{x=c,y=d} \right],$$

and for $\Delta y$:

$$\Delta y = \frac{\lambda f}{2\pi} \left[ \frac{M_1}{M_1 + M_2} \partial_y \varphi(x,y)|_{x=a,y=b} \right. \quad (36)$$
$$\left. + \frac{M_2}{M_1 + M_2} \partial_y \varphi(x,y)|_{x=c,y=d} \right].$$

Introducing the operator $\vec{\nabla}_{P_i} = \left(\hat{\imath}\frac{\partial}{\partial x}, \hat{\jmath}\frac{\partial}{\partial y}\right)\Big|_{P_i=(x_i,y_i)}$, we get:

$$\vec{\Delta} = (\Delta x, \Delta y) = \frac{\lambda f}{2\pi} \left[ \frac{M_1}{M_1 + M_2} \vec{\nabla}_{P_1} \varphi(x,y) \right. \quad (37)$$
$$\left. + \frac{M_2}{M_1 + M_2} \vec{\nabla}_{P_2} \varphi(x,y) \right],$$

which is the averaged sum of the two centroids, corresponding to the results of each individual aperture. We can generalize for an arbitrary number, $N$, of non-overlapping square apertures:

$$\vec{\Delta} = (\Delta x, \Delta y) = \frac{\lambda f}{2\pi} \sum_{i=1}^{N} \frac{M_i}{\sum_{i=1}^{N} M_i} \vec{\nabla}_{P_i} \varphi(x,y). \quad (38)$$

If all the apertures transmit the same energy, the first factor inside the summatory is simply $1/N$, which leads to:

$$\vec{\Delta} = (\Delta x, \Delta y) = \frac{\lambda f}{2\pi N} \sum_{i=1}^{N} \vec{\nabla}_{P_i} \varphi(x,y). \quad (39)$$

So, for a combination of square apertures, the centroid of the resulting distribution is related to the gradient of the phase at each one of the positions of the apertures. If the wavefront is divided into $N$ square regions (or $N$ pixels), Eq. (39) can be used to recover the gradient of the phase at each region by generating N different distributions of squares. This is exactly what we do when we illuminate the object with Hadamard patterns, as each pattern can be viewed as a different combination of squares that either transmit or block the light.

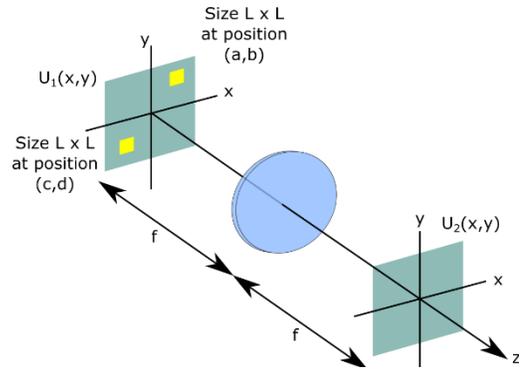

Fig. S5. Masking the wavefront with multiple square masks at the same time. Now, two regions of the wavefront will be transmitted through the system. We will see how the centroid of the distribution in the second plane is related to the phase of the wavefront in the first plane.

## References

1. J. D. Gaskill, *Linear Systems, Fourier Transforms, and Optics* (Wiley, 1978).